\documentstyle[12pt]{article}
\begin{document}
\begin{center}
{\bf 
Coupling Gravity and the Standard Model: a Conformal Approach}
\end{center}
\vskip .5cm

\begin{center} {\bf Marek Paw\l owski}\footnote{Partially supported by 
Polish Committee for Scientific Researches.}  \\ 
Soltan Institute for Nuclear
Studies, Warsaw, POLAND 
\end{center} 
\bigskip\bigskip

\begin{abstract}

In 1992 Professor Ryszard R\c aczka started to work on formulation of
a version of the Standard Model conformally coupled with gravity.  Being his
student I had got the honour to participate in those researches.  The
conformally invariant Higgs-Free Model was the result of our
investigations \cite{hfm,hfm1,hfm2}.  Ryszard R\c aczka passed away on
26 August 1996.  I present here a short memorial survey of our 
results.

\end {abstract} 

In early nineties the Standard Model described reasonably well
$Z_0$ physics and the Standard Model inspired methods could be
successfully applicable to higher and lower energy phenomena.  Thus it
seemed that the only missing element predicted by the theory but not
observed is the Higgs boson.  We wanted to eliminate it from the game
but simultaneously we wanted to preserve as much of successful
predictions of the standard theory as possible.  The conformally
invariant Higgs-Free Model (HFM) which we have proposed
\cite{hfm,hfm1,hfm2} seemed to fulfil these conditions.

Our idea was based on a simple observation:  a gauge transformation
may change the interpretation of a particular degree of freedom (but
of course it does not eliminate it).  If we connect the Standard Model
(with the Higgs doublet) with a conformally invariant gravity (which,
in contrast to the Einstein case, does not contain a dynamical degree
of freedom connected with the "length" of metric tensor) and if we
connect them in such a way that the local conformal symmetry is a
symmetry of the whole model then we would obtain a possibility to
eliminate the redundant degree of freedom in the Higgs sector and to
reincarnate the missing degree of freedom in gravitational sector
simultaneously.

Such a classical model is to be quantized.  As usually in the case of
gauge theories, only the dynamical degrees of freedom are a subject of
quantization in the first step of the construction.  The whole gauge
symmetry can be restored afterwards what is transparent in the path
integral formulation.  Here the Fadeev-Popov trick is a useful tool.
As we have shown \cite{hfm1} the modified Fadeev-Popov method can be
applied in our theory.

Let us describe some details of the model.  The lagrangian of our
Higgs-Free Model reads

$$
L = [L_{SM^c}+ L_{\Phi/g}+ L_{{grav}} ] \sqrt{-g} 
\eqno (1)
$$
where $L_{SM^c}$ is an ordinary Standard Model lagrangian with the 
Higgs doublet $\Phi$ without the Higgs mass term however.
Instead there is a Higgs-gravity interaction term 
$$
L_{\Phi/g}=\beta\partial_\mu|\Phi|\partial^\mu|\Phi|
- {1\over 6}(1+\beta)R\Phi^{\dagger}\Phi, 
\eqno (2)
$$
which contains a Penrose inspired coupling and a $\beta$  proportional 
conformally invariant part. The last one is not very aesthetical but 
it assures the proper ratio of electroweak and gravitational couplings 
at the classical level already. 
Finally a conformally invariant pure gravitational part is given by

$$
L_{{grav}} = -\rho C^2, \hskip1cm\rho\geq 0, 
\eqno (3)
$$
where $C_{\alpha\beta\gamma}^\delta$ is the Weyl tensor.

As was said the essential feature of the model is the local conformal
invariance.  It means that simultaneous rescaling of all fields
(including the field of metric tensor) with a common, arbitrary,
space--time dependent factor $\Omega(x)$ taken with a proper power for
each field (the conformal weight) will leave the Lagrangian (1)
unaffected.  The symmetry has a clear and obvious physical
interpretation.  It changes in every point of the space--time all
dimensional quantities (lengths, masses, energy levels, etc.)  leaving
theirs ratios unchanged.  It reflexes the deep truth of the nature
that nothing except the numbers has an independent physical meaning.
In the conventional approach we define the length scale in such a way
that elementary particle masses are the same for all times and in all
places.  This will be the case when we rescale all fields with the
$x$--dependent conformal factor $\Omega(x)$ in such a manner that the
length of the rescaled scalar field doublet is fixed that is

$$
\tilde\Phi^{\dagger}\tilde\Phi={v^2 \over 2}=const. 
\eqno (4)
$$

This is the phenomenon which was mention at the beginning:  we can
choose the gauge condition in such a way that the Higgs field
disappear but simultaneously we obtain the Einstein like term in the
gravity sector

$$
- {1\over 6}(1+\beta){v^2 \over 2}R. 
\eqno (5)
$$

Obviously we can choose other conditions and obtain other
interpretations of degrees of freedom.

If we take $\beta$ big enough and negative then the term (5) can  be 
fixed to the ordinary Einstein form.  We have check the properties of 
the gravitational sector of our model.  If we impose the condition (4)
 then the variation of (1) with respect to the metric leads to the 
 classical equation of motion which in the empty case $T_{\mu\nu}=0$ 
 is satisfied by all solutions of an empty space Einstein equation 
 with a properly chosen cosmological constant $\Lambda_c$:  
$$
\Lambda_c={3\over2(1+\beta)}\lambda v^2.  
\eqno(6)
$$
($\lambda$ is a quartic Higgs coupling constant of $L_{SM^c}$.) This 
important result was confirmed by other authors \cite{hehl}.  Since 
in our formalism we do not use the spontaneous symmetry breaking 
mechanism (SSB) for mass generation (we will come to this point below) 
the self-interaction term $\lambda(\Phi^\dagger\Phi)^2$ can be set to 
zero by setting $\lambda=0$.  In this case the cosmological  constant 
(6) obtained in our formalism is also zero in agreement with 
experiments and the conviction of Einstein and many others.

The condition (4) together with the unitary gauge fixing of 
$SU(2)_{L}\times U(1)$ gauge group, reduce the Higgs doublet  to 
the form
$$
\Phi^{{scaled}}={1\over\sqrt{2}}{0\choose v}, \hskip 1cm v>0 
\eqno (7)
$$
Inserting this condition into (1) we produce the tree level  mass 
terms for leptons, quarks and vector bosons associated  with 
$SU(2)_L$ gauge group.

The fermion--vector boson interactions in our model are the same as 
in SM.  Hence analogously as in the case of conventional formulation 
of SM one can deduce the tree level relation between $v$ and $G_F$ -- 
the four--fermion coupling constant of $\beta$--decay:  
$$
v^2=(2G_F)^{-1}\rightarrow v=246GeV.  
\eqno (8)
$$ 
(We have used the standard decomposition $g^{\mu\nu}\sqrt{-g}= 
\eta^{\mu\nu}+\sqrt{2}\kappa h^{\mu\nu}$ which reduced the tree level 
problem for the matter fields to the ordinary flat case task.)

The resulting expressions for masses of physical particles are
identical as in the conventional SM and we see that the Higgs
mechanism and SSB is not indispensable for the fermion and vector
mesons mass generation!.

Several words should be written about the renormalizability.  The
lagrangian (1) contains three parts.  Two of them - the Standard Model
and the $R^2$ gravity - seem to be separately renormalizable.  The
nonpolynomial term of the Higgs-gravity interaction part (2) spoils
this property.  The $\beta$ proportional terms are our unwonted
element of the model.  They were introduced in order to square the
strengths of (quantum) weak interactions and the (classical) gravity.
If there would be found a more subtle mechanism of coordination these
quantum and classical effects then the nonpolynomial term would be
redundant.

Despite the mentioned above problems we were able to derive definite
dynamical predictions of our model.  These are electro-weak
predictions for the present accelerator experiments.

It is natural and reasonable for this purpose to ignore gravity and
consider a flat space-time approximation of the model.  In the limit
of flat space-time our model represents a massive vector boson (MVB)
theory for electroweak and strong interactions, which is
perturbatively nonrenormalizable.  The perturbative
nonrenormalizability means that we are not able to improve the
accuracy of predictions by inclusion of more and more perturbation
orders but it does not mean that we are not able to derive predictions
at all.  We can obtain definite predictions at one loop if we
introduce some ultraviolet (UV) cutoff $\Lambda$ and if we consider
processes with energy scale $E$ below this cutoff.  We have shown that
the cutoff $\Lambda$ is closely connected with the Higgs mass $m_H$
appearing in the Standard Model.  From this point of view Higgs mass
is nothing else as the UV cutoff which assures that the truncated
perturbation series is meaningful.

The introduction of $\Lambda$ makes all predictions
$\Lambda$-dependent.  In order to obtain the independent predictions
we have to select some reference observable $R_0(\Lambda)$ which is
measured with the best accuracy in the present EW experiments.  It
will replace the unknown variable $\Lambda$ in the expressions for the
other physical quantities $R_i$.  Choosing the total width of
$Z$-meson $\Gamma_Z$ as the reference quantity we have calculated one
loop predictions for all interesting observables measured at $Z_0$
peak \cite{hfm2}.  It is nothing unexpected that the results almost
coincide with SM predictions and equally well (or poorly) describes
the present data.  One can expect that even more accurate measurements
taken at the single energy point would not be able to discriminate
between the Standard Model and a class of effective models with
massive vector mesons - as I said the Higgs mass plays the role of UV
cutoff and {\sl vice versa}.

One can distinguish these models finding the Higgs boson directly (of
course!).  But we have proposed also another way.

Predictions of an effective model can be in principle calculated for
experiments performed in various energy regions.  These predictions
would depend on the cutoff.  And inversely:  the value of the
appropriate cutoff derived from experimental data collected in
different energy regions can be energy dependent in principle.  The
cutoff is an artificial element which we introduce in order to cover
an incompleteness of the model or imperfection of our calculational
methods.  We try to hide our ignorance in a simplest way:  we
introduce one additional parameter $\Lambda$.  We hope that this
parameter can be the same for a class of similar phenomena.  It would
be nice to have a universal cutoff valid for all phenomena below some
energy scale but in principle it needs not to be the case.  Thus we
have to admit that the cutoff is energy dependent.

We have already mentioned, that the UV cutoff $\Lambda$ is closely
connected with the Higgs mass of SM.  But of course the value of
physical Higgs mass derived from various sets of experiments should be
the same - in contrast to the supposed energy dependence of the cutoff
$\Lambda$.  This is the difference which makes a room for comparison.

In practice we need predictions for the Higgs mass (or the cutoff
$\Lambda$) derived independently from two separate energy regions.
One of them can be of course the $Z_0$ peak.  The second can be
provided for example by 10-20GeV $e^+e^-$ colliders of luminosity high
enough ($\sim$10$^{34}cm^{-2}s^{-1}$ \cite{hfm3}).  We have estimated
that the necessary sample of produced tau pair should be of order of
$\sim10^8$.  Then the observed sensitivity to the value of Higgs mass
will be of order of 100GeV and the predicted Higgs mass could be
compared with the value of $m_H$ derived from $Z_0$-peak data.  We
have to stress that one can relax from the HFM or MVB model
inspiration and regard the proposed test as a selfconsistency check of
the SM itself.

Concluding this memorial overview I'd like to stress that the HFM
provides a novel and original approach to the problem of unified
description of fundamental interactions.  It glues the SM and the
conformal gravity in a conformally invariant way.  As a result we
obtain Einstein-like gravity coupled in the usual way with the version
of SM without the Higgs boson.  All experimentally confirmed results
of Einstein theory and the Standard Model can be naturally reproduced
within the present model.  New phenomena can appear however.  We have
proposed an experimental framework for testing them. The physical Higgs 
boson is absent from the proposed theory!

\end{document}